\documentclass[aps,prl,twocolumn,showpacs,groupedaddress]{revtex4}  
\usepackage{graphicx}  
\usepackage{dcolumn}   
\usepackage{bm}        
\usepackage{amssymb}   
\usepackage{sidecap}   

\newcommand{\etana}{\mbox{$e\tau_h$}}
\newcommand{\mtana}{\mbox{$\mu\tau_h$}}
\newcommand{\emana}{\mbox{$e\mu$}}
\newcommand{\ttype}{\mbox{$\tau$-type}}
\newcommand{\ppbar}{\mbox{$p\overline{p}$}}
\newcommand{\pbs}{\mbox{$\rm{pb}^{-1}$}}
\newcommand{\fbs}{\mbox{$\rm{fb}^{-1}$}}

\newcommand{\ttbar}{\mbox{$t \overline{t} $}}
\newcommand{\tanb}{\mbox{$\tan \beta$}}
\newcommand{\Gev}{${\rm{GeV}}$}
\newcommand{\Tev}{${\rm{TeV}}$}
\newcommand{\Pet}{\mbox{\ensuremath{\not \!\! P_T}}}
\newcommand{\met}{\mbox{\ensuremath{\not \!\! E_T}}}
\newcommand{\metx}{\mbox{\ensuremath{\not \!\! E_x}}}
\newcommand{\mety}{\mbox{\ensuremath{\not \!\! E_y}}}
\newcommand{\Gc}{${\rm{GeV}}$}
\def \mvis {\ensuremath{ M_{\rm{vis}} }}
\def \lep  {\ensuremath{ \ell }}

\begin{document}

\hspace{5.2in} \mbox{Fermilab-Pub-08-132-E}

\title{Search for Higgs bosons decaying to tau pairs in \mbox{$p\overline{p}$} collisions
with the D0 detector}
%
\author{V.M.~Abazov$^{36}$}
\author{B.~Abbott$^{75}$}
\author{M.~Abolins$^{65}$}
\author{B.S.~Acharya$^{29}$}
\author{M.~Adams$^{51}$}
\author{T.~Adams$^{49}$}
\author{E.~Aguilo$^{6}$}
\author{S.H.~Ahn$^{31}$}
\author{M.~Ahsan$^{59}$}
\author{G.D.~Alexeev$^{36}$}
% ...

\affiliation{\vspace{0.1 in}(The D\O\ Collaboration)\vspace{0.1 in}}
% ...
\affiliation{$^{6}$University of Alberta, Edmonton, Alberta, Canada,
                Simon Fraser University, Burnaby, British Columbia, Canada,
                York University, Toronto, Ontario, Canada, and
                McGill University, Montreal, Quebec, Canada}
% ...
\affiliation{$^{29}$Tata Institute of Fundamental Research, Mumbai, India}
% ...
\affiliation{$^{31}$Korea Detector Laboratory, Korea University, Seoul, Korea}
% ...
\affiliation{$^{36}$Joint Institute for Nuclear Research, Dubna, Russia}
% ...
\affiliation{$^{49}$Florida State University, Tallahassee, Florida 32306, USA}
% ...
\affiliation{$^{51}$University of Illinois at Chicago,
                Chicago, Illinois 60607, USA}
% ...
\affiliation{$^{59}$Kansas State University, Manhattan, Kansas 66506, USA}
% ...
\affiliation{$^{65}$Michigan State University,
                East Lansing, Michigan 48824, USA}
% ...
\affiliation{$^{75}$University of Oklahoma, Norman, Oklahoma 73019, USA}
% ...

\date{May 16, 2008}

\begin{abstract}
We present a search for the production of neutral Higgs bosons $\phi$~decaying
into $\tau^+\tau^-$ final states in \ppbar\ collisions
at a center-of-mass energy of 1.96~TeV.
The data, corresponding to an integrated luminosity of approximately 1~\fbs,
were collected by the D0 experiment at the Fermilab Tevatron Collider.
Limits on the production cross section times branching ratio are set. 
The results are interpreted in the minimal supersymmetric standard model
yielding limits that are the most
stringent to date at hadron colliders.

\end{abstract}

\maketitle 

Higgs bosons are an essential ingredient of electroweak symmetry breaking in the standard model (SM).
A search for Higgs bosons (denoted as $\phi$) decaying to tau leptons is of particular interest in models with more than one Higgs doublet,
where production rates for $p\bar{p}\rightarrow\phi\rightarrow\tau^+\tau^-$ can potentially be large enough for observation at the
Fermilab Tevatron Collider. This situation is realized in
the minimal supersymmetric standard model (MSSM)~\cite{MSSM},
which contains two complex Higgs doublets,
leading to two neutral CP-even ($h,H$), one CP-odd ($A$), and a pair of charged ($H^{\pm}$) Higgs bosons. At tree level, the Higgs
sector of the MSSM is fully specified by two parameters, generally chosen to be $M_A$, the mass of the CP-odd Higgs boson,
and \tanb, the ratio of the vacuum expectation values of the two Higgs doublets.
Dependence on other MSSM parameters enters through radiative corrections.
At large \tanb, the coupling of the neutral
Higgs bosons to down-type quarks and charged leptons is strongly enhanced,
leading to sizable cross sections. The Higgs bosons will
decay predominantly into third generation fermions.

Searches for neutral MSSM Higgs bosons have been conducted at LEP \cite{LEP-limit} and at the
Tevatron \cite{dzero-bb,CDF-tautau,dzero-p14tautau}.
These Tevatron searches used between 260~\pbs\ and 350~\pbs\ of collider data.
In this Letter a search for $\phi \to \tau^+ \tau^-$~with about 1~\fbs~\cite{lumi} of data is presented.
At least one of the tau leptons is required to decay leptonically,
leading to final states containing $e  \tau_h$, $\mu  \tau_h$ and $e \mu$,
where $\tau_h$ represents a hadronically decaying tau lepton. 
The data were collected at the Tevatron with the D0 detector
between 2002 and 2006 at a $p\overline{p}$~center-of-mass
energy $\sqrt{s} = 1.96$~TeV.
A description of the D0\ detector can be found in Ref.~\cite{d0det}.

Signal and SM background processes are modeled using the {\sc pythia}~6.329~\cite{pythia} Monte Carlo (MC) generator,
followed by a {\sc geant}-based~\cite{geant} simulation of the D0 detector.
The signal events are produced with the width of the SM Higgs boson.
All background processes, apart from multijet production and $W$ boson production, are normalized using
cross sections calculated  at next-to-leading order (NLO) and next-to-NLO
(for $Z$ boson and Drell-Yan production) based on the CTEQ6.1~\cite{cteq} parton distribution functions (PDF).

The normalization and shape of background contributions from multijet production, where jets are misidentified as leptons,
are estimated from the data by using same charge $e$ and $\tau_h$ candidate events (\etana\ channel) or by selecting
background samples by inverting lepton identification criteria (\mtana\ and $e \mu$ channels).
These samples are normalized to the data at an early stage of the selection in a region of phase space dominated by multijet production.
The multijet background estimation in the \mtana\ and \etana\ channels was checked by using an independent method to estimate
the background: in the \mtana\ channel same charge $\mu\tau_h$~events were used and in \etana\ channel the
multijet background was estimated from measurements in data
of the probability to mis-reconstruct electrons from jets. The differences between the estimates
were used to set the systematic uncertainty on the multijet production.
The normalization of the background from $W$~boson production is obtained from data in a sample dominated by $W~\mbox{boson} + \mbox{jet}$
events.

Electrons are selected using
their characteristic energy deposits,
including the transverse and longitudinal shower profile in the
electromagnetic (EM) calorimeter.
To reject photons, a reconstructed track is required to point to the energy cluster.
Further rejection against background is achieved by using a
likelihood discriminant.
Muons are selected using reconstructed tracks in the central tracking detector in
combination with patterns of hits in the muon detector.
Muons are required to be isolated in the calorimeter and the tracker~\cite{hpp}.
Reconstruction efficiencies for both leptons are measured in data using
$Z/\gamma^*\rightarrow \mu^+\mu^-,e^+e^-$ events.

A hadronically decaying tau lepton is characterized by a narrow isolated jet with low track multiplicity~\cite{d0-z-tautau}.
Three \ttype s are distinguished:
\ttype\ 1 is a single track with energy deposited in the hadronic calorimeter ($\pi^\pm$-like);
\ttype\ 2 is a single track with energy deposited in the hadronic and the electromagnetic calorimeters ($\rho^\pm$-like);
\ttype\ 3 is three tracks with an invariant mass below 1.7~\Gev, with energy deposited in the calorimeter.

A set of neural networks, $NN_{\tau}$, one for each \ttype,
has been trained to separate hadronic tau decays from jets
using $Z/\gamma^*\rightarrow\tau^+\tau^-$~MC as signal and
multijet data as background.
The selections on the neural networks retain $66\%$~of the $Z/\gamma^*\rightarrow\tau^+\tau^-$~events, while
rejecting $98\%$~of the multijet background.
In addition, a neural network has been trained
with electron MC events as background
to separate \ttype~2 hadronic
tau candidates from electrons ($NN_e$). 

The signal is characterized by two leptons, missing transverse energy $\met$~and
as an enhancement above the background in the visible mass
\mbox{$\mvis=\sqrt{(P_{\tau_1}\ + \ P_{\tau_2} \ + \ \Pet)^2}$}, 
calculated using the four-vectors of the visible tau decay products $P_{\tau_{1,2}}$ and
of the missing momentum $\Pet = (\met,\metx,\mety, 0)$. The components $\metx$ and $\mety$ 
of $\met$~are computed from calorimeter cells and the momentum of muons,
and corrected for the energy response of electrons, taus and jets.
The four-vectors of the hadronic taus are calculated using the calorimeter
for \ttype s 2 and 3 and
the central tracking system for \ttype~1.

\begin{SCfigure*}
\centering
\includegraphics[height=40mm]{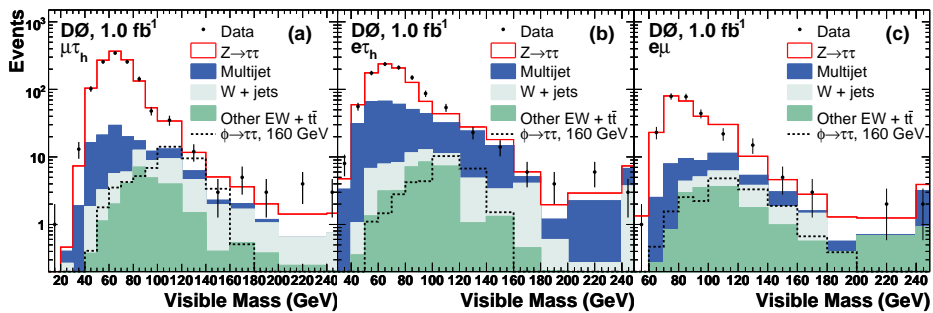}
\caption[]{The distribution of the visible mass $\mvis$ for (a) \mtana, (b) \etana\ and
(c) \emana~channels. The Higgs boson signal is normalized to a cross section of $3$~pb. 
The highest bin includes the overflow.}
\label{f:vis-mass}
\end{SCfigure*}

In the \etana\ and \mtana\ channels, an isolated lepton ($e, \mu$) with transverse momentum above 15~\Gc\
and an isolated hadronic tau with transverse momentum above $16.5$~\Gc\ ($22$~\Gc\ for \ttype\ 3)
are required. The pseudorapidity $|\eta|$~is less than $2$ for muons and hadronic taus and
$2.5$ for electrons.
In addition to the background from $Z/\gamma^*\rightarrow\tau^+\tau^-$ production,
a $W(\rightarrow\lep\nu) +$~jet event can be misidentified as a high-mass di-tau event if the jet
is misidentified as a hadronic tau decay.
The transverse mass,  $M^{e/\mu}_T=[2p^{e/\mu}_{T}\met(1-\cos \Delta\varphi)]^{\frac{1}{2}}$,
is required to be less than 40~\Gc\ for the \mtana\ and 50~\Gc\ for the \etana\ channel.
Here, $\Delta\varphi$ is the azimuthal angle between the lepton and $\met$.
In addition, a selection is made in the $\Delta\varphi(e/\mu,\met) - \Delta\varphi(\tau,\met)$ plane, such that
$\Delta\varphi(e/\mu,\met) < 3.5 - \Delta\varphi(\tau,\met)$~if $\Delta\varphi(\tau,\met)<2.9$~or $\Delta\varphi(e/\mu,\met)<0.6$~otherwise.
This selection removes events where the missing transverse energy is in the
hemisphere opposite to the muon and the tau candidate. 
Due to the larger multijet background in the \etana\ channel the azimuthal angle between the electron and tau,
$\Delta\varphi(e,\tau)$,~is required to be greater than 1.6.

The \etana\ channel has a significant background from $Z/\gamma^*\rightarrow e^+e^-$ production, where an electron is mis-reconstructed as a
tau candidate. To remove these events, the tau candidates in the \etana\ channel are required to be outside of
the region $1.05<|\eta|<1.55$, where there is limited EM calorimeter coverage
and are required to have less than $90\%$ of their energy deposited in the EM calorimeter.
Finally, \ttype~2 candidates are required to have $NN_e > 0.8$, which
rejects $92\%$~of the $Z/\gamma^*\rightarrow e^+e^-$~events, while retaining
$83\%$~of the $Z/\gamma^*\rightarrow\tau^+\tau^-$~events.

We select one muon with \mbox{$p_T >$ 10~\Gc} and
one electron with \mbox{$p_T >$ 12~\Gc} in the \emana\ channel.
Multijet and $W$ boson production are suppressed by requiring the
invariant mass of the electron-muon pair
to be above 20~\Gc\ and $\met + p_{T}^{\mu} + p_{T}^e > 65$~\Gc.
Background from $W$+jet events can be reduced using the transverse mass
by requiring that either $M_T^e<10$~GeV or $M_T^\mu<10$~GeV.
Furthermore, the minimum angle between the leptons and the \met\ vector, $\min[\Delta\varphi(e,\met),\Delta\varphi(\mu,\met)]$,
has to be smaller than $0.3$.
Contributions from \ttbar\ background are suppressed by rejecting events
where the scalar sum of the transverse momenta
of all jets in the event is greater than 70~\Gc.

\begin{table}
\caption{Expected number of events for backgrounds, number of events observed in the data and efficiency for a signal with \mbox{$M_\phi=160$~GeV} for the three channels. The uncertainties are statistical. \label{t:evt_numbers}}
\begin{tabular}{l@{\hspace{0.45cm}}r@{$\,\pm \,$}l@{\hspace{0.2cm}}r@{$\,\pm \,$}l@{\hspace{0.2cm}}r@{$\,\pm \,$}l}
\hline
Channel         & \multicolumn{2}{c}{$e \tau_h$}    & \multicolumn{2}{c}{$\mu \tau_h$}  & \multicolumn{2}{c}{$e \mu$} \\
\hline
$Z / \gamma^* \to \tau^+ \tau^-$     &  581 & 5    & 1130  & 7     & 212  & 3   \\
Multijet                         &  332 & 20   &  86   & 4     &  29  & 1 \\ 
$W \to e \nu, \mu \nu, \tau \nu$ &   42 & 5    &  32   & 4     &   9  & 2 \\
$Z / \gamma^* \to e^+e^-, \mu^+ \mu^-$   &   31 & 2    &  19   & 1     &  12  & 1 \\
Diboson + $\ttbar$              &  3.0 & 0.1  &  7.0  & 0.4   &  6.1 & 0.1 \\ \hline
Total expected                   &  989 & 23   & 1274  & 9     &  269 & 3   \\ \hline
Data             &   \multicolumn{2}{c}{1034}      &  \multicolumn{2}{c}{1231}        & \multicolumn{2}{c}{274} \\  \hline
Efficiency (\%)                    & 1.04 & 0.03 & 1.46  & 0.04  &  0.57  & 0.03 \\ \hline
\end{tabular}

\end{table}

The number of events observed in the data and expected from the various SM processes
show good agreement (Table~\ref{t:evt_numbers}).
The number of background and signal events depend on numerous measurements that introduce a systematic uncertainty:
integrated luminosity (6.1\%), trigger efficiency (3\%--4\%), lepton identification and reconstruction efficiencies (2\%--10\%),
jet and tau energy calibration (2\%--3\%), PDF uncertainty (4\%), the uncertainty on the $Z/\gamma^*$ production cross section (5\%),
normalization of the $W$~boson background (6\%--15\%),
and modeling of multijet background (4\%--40\%).
All except the last one are correlated among the three final states.
Most of the uncertainties affect only the overall acceptance for the signal and backgrounds. However,
uncertainties on the energy scale
and electron trigger efficiencies modify the shape of the visible mass distribution
(Fig.~\ref{f:vis-mass}). These uncertainties are therefore parameterized
as a function of $\mvis$.

\begin{figure}[ht!!!]
\begin{center}
\includegraphics[height=40mm]{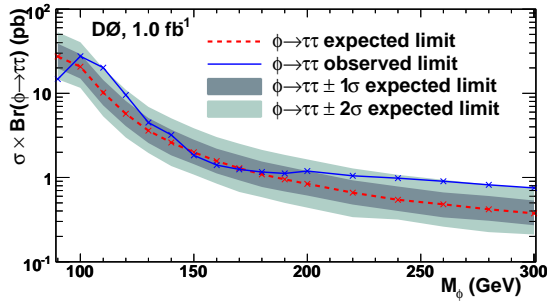}
\caption[]{
Expected and observed 95\% CL upper limits on the cross section times branching ratio for
$\phi \to \tau^+\tau^-$~production
as a function of $M_{\phi}$ assuming the SM width of the Higgs boson.
The $\pm 1,2$~standard deviation bands on the expected limit are
also shown.
}
\label{f:xs-limit}
\end{center}
\end{figure}

We extract upper limits on the production cross section times branching ratio as a function of
Higgs boson mass $M_\phi$. In order to maximize the sensitivity (median expected limit), 
the event samples of the \etana\ and \mtana\ channels are separated by \ttype\ to exploit the different signal-to-background ratios.
Furthermore the differences in shape between signal and background are exploited by using the full
$\mvis$~spectrum in the limit calculation (Fig.~\ref{f:vis-mass}).
The limits are calculated by utilizing a likelihood-fitter~\cite{wadelimit} that uses a 
log-likelihood ratio test statistic method.
The confidence level, $CL_s$, is defined as $CL_s = CL_{s+b} / CL_b$,
where $CL_{s+b}$ and $CL_b$~are the confidence levels in the signal-plus-background and background-only hypotheses respectively.
The expected and observed limits are calculated by scaling the signal until $1-CL_s$ reaches $0.95$.
The resulting cross section limits are shown in Fig.~\ref{f:xs-limit}.
The difference between the observed and expected limits
at high masses is slightly above two standard deviations.
It is mainly caused by a data excess in the $\mtana$ channel
above $\mvis$~of $160$~GeV.
A large number of kinematic distributions
were studied for this sample and the data are consistent with both background
and signal shapes. Due to the $\mvis$~resolution these events
affect the limit over a wide range of masses.

\begin{figure}
\begin{center}
\includegraphics[height=40mm]{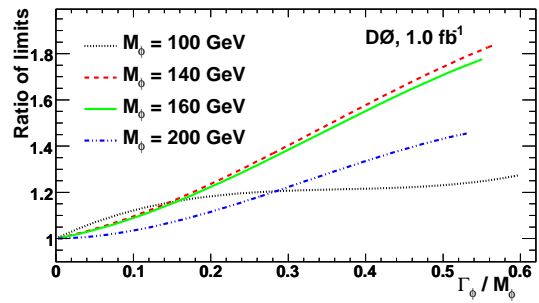}
\caption[] {
Ratio of expected cross section limits using a Higgs boson with
non-SM width to those calculated with
a Higgs boson with SM width, as a function of $\Gamma_{\phi} / M_{\phi}$.
}
\label{f:widelimit}
\end{center}
\end{figure}

The limits in Fig.~\ref{f:xs-limit} assume a Higgs boson with SM width,
which is negligible compared to the experimental resolution on $\mvis$.
In models such as the MSSM the Higgs boson width can become substantially
larger than the value in the SM. This was simulated by multiplying a
relativistic Breit-Wigner ($BW$) function
with the cross section from {\sc feynhiggs}~\cite{feynhiggs}
for masses $M>80$~GeV to obtain the differential cross section
for a wide Higgs boson as a function of mass:
\begin{equation}
\frac{d\sigma}{dM} = \sigma(M, \tan\beta, \Gamma_{\phi}=0) \times BW(M,M_{\phi},\Gamma_{\phi}).
\label{e-wideh}
\end{equation}
This differential cross section was used to build a signal template of the $\mvis$~distribution
for a Higgs boson of mass $M_{\phi}$ and width $\Gamma_{\phi}$.
The limit calculation procedure
was then repeated with templates corresponding to various values of $\Gamma_{\phi}$. The
ratio of the expected cross section limit for a wide Higgs boson to the limit for a Higgs boson
with SM width as a function of $\Gamma_{\phi} / M_{\phi}$
is shown in Fig.~\ref{f:widelimit}. This result can be used to correct the
cross section limit for a Higgs boson with SM width (Fig.~\ref{f:xs-limit}) for a non SM width
in a model independent way.

In the MSSM, the masses and couplings of the Higgs bosons depend, in addition to \tanb\ and $M_A$, on the MSSM parameters through radiative corrections. In a constrained model, where unification of the SU(2) and U(1) gaugino masses is assumed, the most relevant parameters are the mixing parameter $X_t$, the Higgs mass parameter $\mu$, the gaugino mass term $M_2$, the gluino mass $m_g$, and a common scalar mass $M_{\rm SUSY}$.  
Limits on $\tan \beta$ as a function of $M_A$ are derived for two scenarios assuming a CP-conserving Higgs sector~\cite{mssm-benchmark}: the $m_h^{\rm max}$ scenario~\footnote{$M_{\rm SUSY}$= 1~\Tev, $X_t$ = 2~\Tev, $M_2$ = 0.2~\Tev,
$\mu$ = $+$0.2~\Tev, and $m_g$ = 0.8~\Tev.} and the no-mixing
scenario~\footnote{$M_{\rm SUSY}$= 2~\Tev, $X_t$ = 0~\Tev, $M_2$ = 0.2~\Tev, $\mu$ = $+$0.2~\Tev, and $m_g$ = 1.6~\Tev.} with $\mu=+0.2$~\Tev.
The $\mu<0$~case is not considered as it is currently disfavored~\cite{Ellis:2007fu}.
The production cross sections, widths, and branching ratios for the Higgs bosons are calculated over the mass range from 90~\Gev\ to 300~\Gev\ using the {\sc feynhiggs} program \cite{feynhiggs}.
\begin{figure}[htb]
\begin{center}
\includegraphics[height=40mm]{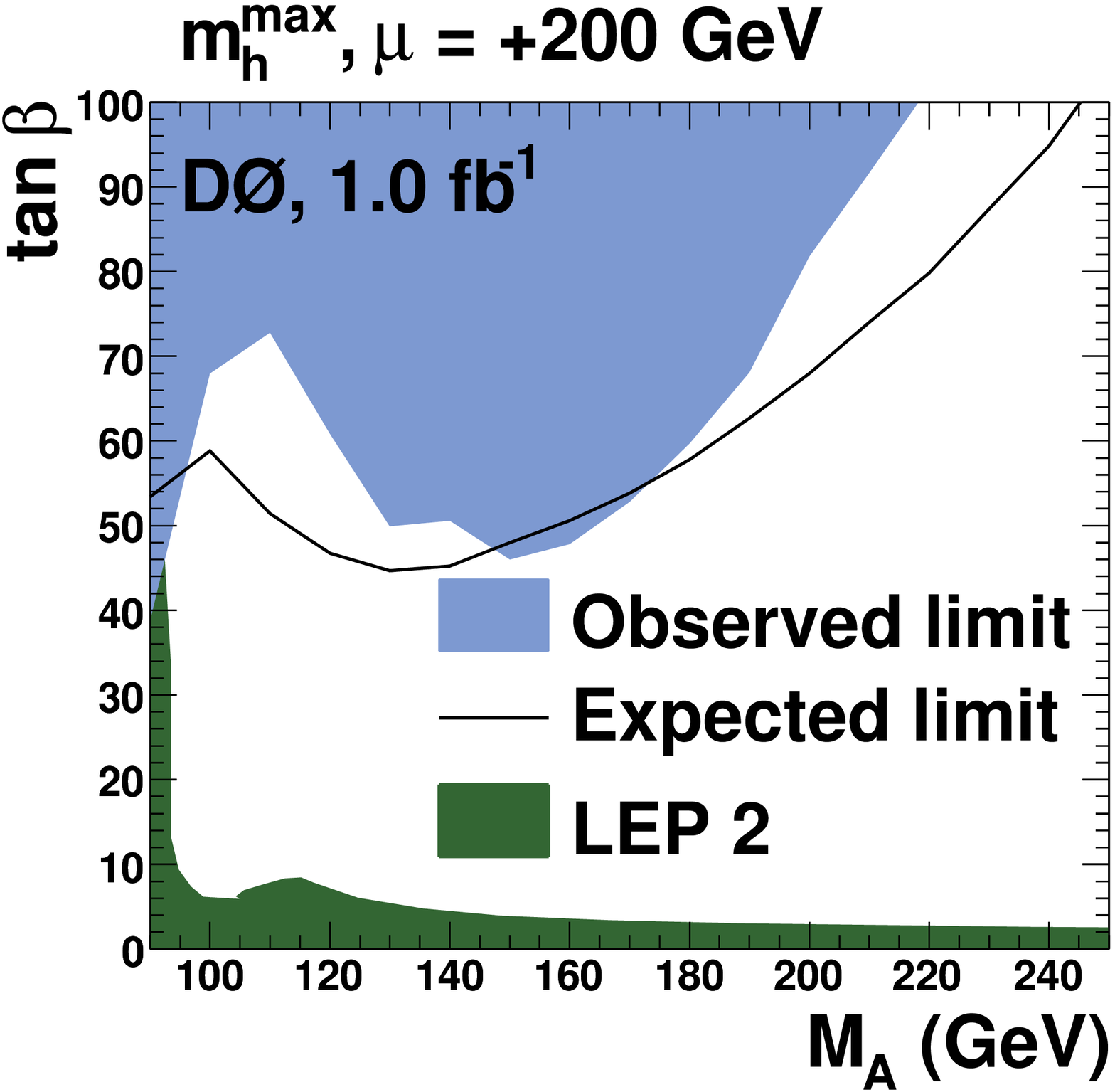}
\includegraphics[height=40mm]{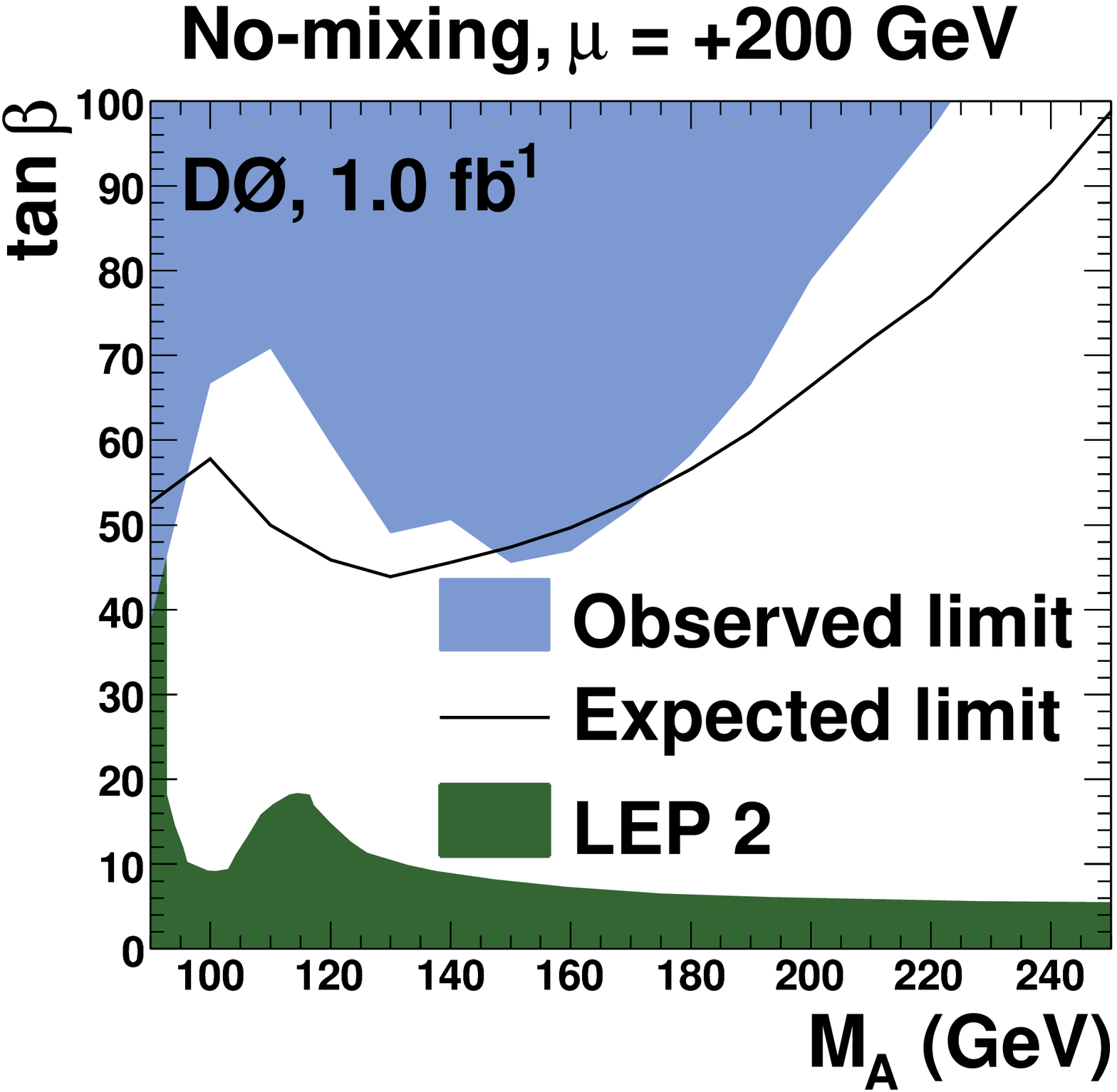}

\caption[]{Region in the ($M_A, \tanb$) plane that is excluded at 95\% CL for the $m_h^{\rm max}$ and the no-mixing scenario \mbox{($m_{t}=172.6$~GeV~\cite{mtop})}. 
Also shown is the excluded region from LEP~\cite{LEP-limit}.}
\label{f:tanb-excl}
\end{center}
\end{figure}
In these scenarios $\Gamma_A / M_A < 0.1$~for
$M_A < 200$~GeV.
The effect of the Higgs boson width is therefore small.
For large \tanb, the $A$ boson is nearly degenerate in mass with either the $h$ or the $H$ boson,
and their production cross sections ($gg\rightarrow\phi$,~$b\overline{b}\rightarrow\phi$)
are added.

Fig.~\ref{f:tanb-excl} shows the results
interpreted in the MSSM scenarios considered in the Letter.
We reach a sensitivity of around $\tan\beta=50$~for $M_A$~below
180~\Gev.
The result represents the most stringent limit on the production of
neutral MSSM Higgs bosons at hadron colliders.

%
We thank the staffs at Fermilab and collaborating institutions, 
and acknowledge support from the 
DOE and NSF (USA);
CEA and CNRS/IN2P3 (France);
FASI, Rosatom and RFBR (Russia);
CNPq, FAPERJ, FAPESP and FUNDUNESP (Brazil);
DAE and DST (India);
Colciencias (Colombia);
CONACyT (Mexico);
KRF and KOSEF (Korea);
CONICET and UBACyT (Argentina);
FOM (The Netherlands);
STFC (United Kingdom);
MSMT and GACR (Czech Republic);
CRC Program, CFI, NSERC and WestGrid Project (Canada);
BMBF and DFG (Germany);
SFI (Ireland);
The Swedish Research Council (Sweden);
CAS and CNSF (China);
and the
Alexander von Humboldt Foundation.

\end{document}